# Semiconductor Devices Condition Monitoring Using Harmonics in Inverter Control Variables

Shuyu Ou*, Ariya Sangwongwanich, Subham Sahoo, and Frede Blaabjerg
AAU ENERGY, AALBORG UNIVERSITY
9220 Aalborg East, Denmark
Tel.: +46 / (0) – 767 090 877.
∗E-Mail: so@energy.aau.dk
URL: https://vbn.aau.dk/en/persons/155107

## Acknowledgements

This project is supported by the European Union's Horizon 2020 research and innovation program under the Marie Skłodowska-Curie grant agreement No 955614.

## Abstract

The health status of power semiconductor devices in power converters is important but difficult to monitor. This paper analyzes the relationship between harmonics in inverter control variables and a health precursor (the on-state voltage $V_{on}$ of power semiconductor devices). Based on the analysis, harmonics can estimate $V_{on}$ without adding extra sensing circuits. The method is validated through simulations.

## Introduction

Power electronic-based converters are used in a wide range of applications including electric vehicle powertrains, aerospace power supplies, distributed generation systems, etc. Besides high power density and efficiency, reliability is a main concern of power converters since it affects the system performance and cost. The abnormal states of inverters is the highest (43%) among all the other devices leading to 36% lost productivity in photovoltaic (PV) systems [1]. Power converters also contribute to 13% of failures over different subsystems and 18% of downtime in wind turbines [2].

Predictive maintenance is one of the solutions to improve system reliability and reduce the unscheduled maintenance cost. It can help to optimize maintenance plan for power converter applications based on the health status of the critical components, e.g., the power semiconductor devices [3]. Various condition precursors can be used to infer health status of power semiconductor devices, wherein the on-state voltage ($V_{on}$) is one of them. The on-state voltage $V_{on}$ has several advantages including high sensitivity, high accuracy, easier calibration, and better online measurement capability [4]. It can potentially indicate several types of failure mechanisms: bond wires lift-off/cracking, solder fatigue, and delamination of die attach in a power module [5]-[7].

A general method to monitor the on-state voltage $V_{on}$ is through an extra sensing circuit. Extra devices are added to the inverter output terminal [5], [8], [9] or the gate driver [10]. The monitored devices are multiple power semiconductors in a half-bridge [9], [11] or in a converter [5], [8]. The challenges for the on-state voltage measurement include short measurement time, high switching noises, and low voltage variation caused by degradation. Consequently, adding extra sensing circuits will increase the amount of components as well as cost. Extra components may also introduce additional risk of failure to the overall system.

To avoid adding extra circuits, D. Xiang et al. [12] and F. Yüce et al. [13], [14] exploited harmonics in the control variables as a precursor to monitor power semiconductor devices in a three-phase inverter. However, no general mathematical justification is provided, which limits the generalization of this method to a specific system. As a result, recalibration is repeated for different systems. Nevertheless, the mechanism behind the phenomenon should be analyzed in detail to map harmonics uniformly for health indication in different applications.

Therefore, this paper investigates the mechanism of how the on-state voltage $V_{on}$ adds harmonics to the inverter output voltage. These harmonics propagate within the closed current control loop which means the $V_{on}$ can be observed from several variables in the control loop. This paper selects only one of the variables: the current controller output (also known as the voltage references in the $dq$ frame) to monitor $V_{on}$. The increment in $V_{on}$ consequently leads to an increment in voltage reference harmonics. Finally, the effectiveness of the proposed method and its limitations are analyzed.

An advantage of the modelling method is this method avoids adding measurement circuits to each individual power semiconductor devices in a converter, which saves space and cost. Additionally, the method is based on a general mathematical model so it should be widely applicable without a calibration process. However, there are two main limitations: First, the $V_{on}$ is only one of the sources of harmonics. The impact of the other harmonic sources, e.g., deadtime, also need to be considered. Second, the inverter output current should be sampled with a high-resolution sampling circuit, with which the current control loop can observe the voltage error caused by degradation.

This paper is organized as follows: Section I presents the inverter topology, control loops, and the on-state voltage $V_{on}$ characteristic. The mechanism of $V_{on}$ inducing harmonics in the current control loop is also introduced in this section. Section II presents the modelling method, based on which the harmonics equations are derived. Section III compares the modelling method with simulation. Performance and sensitivity analysis are also investigated. At the end, the last section summarizes this work.

# I. System Configuration and Device Degradation Characteristic

## A. Inverter Topology and Control Loops

A three-phase two-level grid-connected voltage source inverter (VSI) as shown in Fig. 1 is considered in this work since it is widely used in various applications. The DC-Link capacitor $C_{bus}$ and its equivalent internal series resistance $R_C$ are connected to a power source, e.g., PV panels or a front-end DC-DC conversion stage. The DC bus voltage $v_{dc}$ is converted into AC inverter output voltages $v_{an}$, $v_{bn}$, and $v_{cn}$ with the help of power semiconductor devices including six IGBTs ($S_1$-$S_6$) and six diodes ($D_1$-$D_6$). The inverter stage is connected to the grid $v_g$ through a filter and a grid impedance. The filter consists of a filter inductor $L_g$ and its equivalent resistance $R_L$. The grid impedance consists of inductors $L_s$ and resistors $R_s$.

The inverter is controlled by two controllers using sinusoidal pulse width modulation (SPWM) method. An outer bus voltage control loop uses a proportional integral (PI) controller to maintain the bus voltage $v_{dc}$. The bus voltage loop provides a $d$-axis reference current $i_d^*$ to the current controller and sets the $q$-axis reference current $i_q^*$ to zero. The inner current loop controls the output currents $i_a$, $i_b$, and $i_c$ through changing reference voltages $v_{dq}^*$ ($v_d^*$ and $v_q^*$) in the direct-quadrature $dq$-synchronous rotating frame as well as the reference voltages $v_{abc}^*$ ($v_a^*$, $v_b^*$, and $v_c^*$) in the $abc$ frame. The coordinate transformation requires a grid phase angle $\theta$ coming from the phase-locked loop (PLL). The PWM modulator provides gate signals $v_{gate}$ for IGBTs.

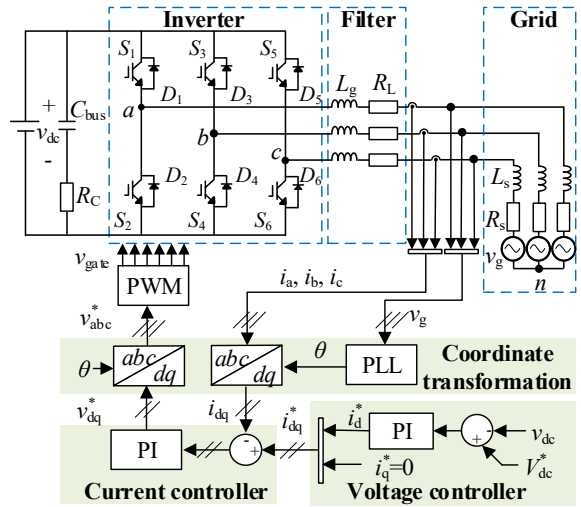

Fig. 1: A three-phase two-level grid-connected inverter controlled by an outer bus voltage loop and an inner current loop.

## B. On-State Voltage $V_{on}$ Characteristic

Power semiconductor devices in an inverter include IGBTs and diodes. These two components have an on-state voltage $V_{on}$ characteristics, as shown in Fig. 2 (a), which can be divided into two equivalent parts: a constant voltage part $V_{on0}$ and a resistive part $R_{on}$, where $I$ is the current flowing through the device, $V_{on}$ can be given by:

$$V_{on} = V_{on0} + R_{on}I \qquad (1)$$

Previous research work have demonstrated that with the aging process, e.g., the on-state voltage $V_{on}$ increases under accelerated power cycling tests, as illustrated in Fig. 2 (b) [4], [5], [9]. The end-of-life criteria $V_{\text{on\_end-of-life}}$ for semiconductor devices is usually set as 5% to 20% increment of the initial voltage $V_{\text{on\_initial}}$ [5]. An increased $\Delta R_{\text{on}}$ can indicate package failures, e.g., bondwire fatigue [6] while $V_{\text{on0}}$ is relatively stable within the whole lifetime. Therefore, this study mainly focuses on the effect of $\Delta R_{\text{on}}$ on voltage reference harmonics.

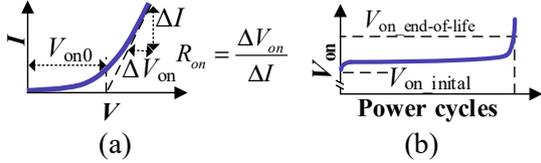

Fig. 2: The on-state voltage $V_{\text{on}}$ characteristics: (a) The transfer characteristic. (b) $V_{\text{on}}$ increases in accelerated power cycling tests [4].

## C. On-State Voltage Von Introduces Harmonics to the Current Loop

In the closed current control loop, the increase in the on-state voltage $V_{\text{on}}$ adds errors (a blue dashed line in Fig. 3) to the inverter output voltages. The current controller responds to the error in the phase current by changing the controller outputs, i.e., the reference voltages $v^*_{dq}$ in the $dq$ frame. Since $v^*_{dq}$ are variables within the controller, using $v^*_{dq}$ avoids adding extra measurement devices. Some other control variables can also monitor the voltage error, e.g., the reference voltages in the $abc$ frame $v^*_{abc}$. However, using $v^*_{abc}$ means implementing frequency analysis to three signals, while using $v^*_{dq}$ only requires two signals to be analyzed, which saves computation power.

In general, the bandwidth of a grid-connected inverter voltage control loop $f_{\text{BV}}$ is lower than 100 Hz, while the bandwidth of the current control loop $f_{\text{BI}}$ is in the range of around one to several kHz. Since the harmonics to be studied are in the range of several times of the fundamental frequency, which is higher than $f_{\text{BV}}$ but lower than $f_{\text{BI}}$, the dynamic of the voltage control loop can be neglected and only the dynamic of the current control loop should be studied.

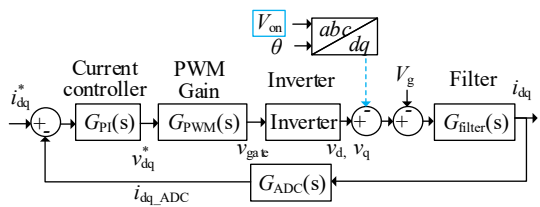

Fig. 3: The on-state voltage $V_{\text{on}}$ adds errors to the current control loop.

According to [15] and [16], the transfer function from the voltage error to the voltage reference $v^*_{dq}$ can be given as in (2).

$$G_{Von}(s) = \frac{v^*_{dq}(s)}{V_{on}(s)}$$
$$= \frac{G_{PI}(s)G_{filter}(s)G_{ADC}(s)}{1+G_{PWM}(s)G_{PI}(s)G_{filter}(s)G_{ADC}(s)}$$
$$G_{PI}(s) = K_{pc} + \frac{K_{ic}}{s}, G_{filter}(s) = \frac{1}{sL_g + R_L},$$
$$G_{ADC}(s) = \frac{1}{0.5T_{sa}s+1}, G_{PWM}(s) = \frac{1}{0.25T_{sa}s+1},$$
(2)

where $G_{Von}$, $G_{PI}$, $G_{filter}$, $G_{ADC}$, and $G_{PWM}$ are transfer functions of the closed-loop errors, the PI current controller, the inductor filter stage, the ADC sampling delay, and the PWM modulator transport delay, respectively. $K_{pc}$ and $K_{ic}$ are the proportional gain and the integral gain of the PI current controller, respectively. $T_{sa}$ is the sampling time.

Two cases of time-domain errors in phase A are shown in Fig. 4, in which switching transients are neglected for simplicity.

- (a) The envelope of the inverter output voltage $v_{an}$ is a two-level waveform, when the on-state voltage of all phase A IGBTs and diodes ($S_1$, $S_2$, $D_1$, and $D_2$) are zero. The upper envelope is at the bus voltage level $v_{dc}$ and the bottom level is zero.
- (b) The non-ideal on-state voltage $V_{\text{on}}$ of $S_1$, $S_2$, $D_1$, and $D_2$ affects the inverter output voltage $v_{an}$.

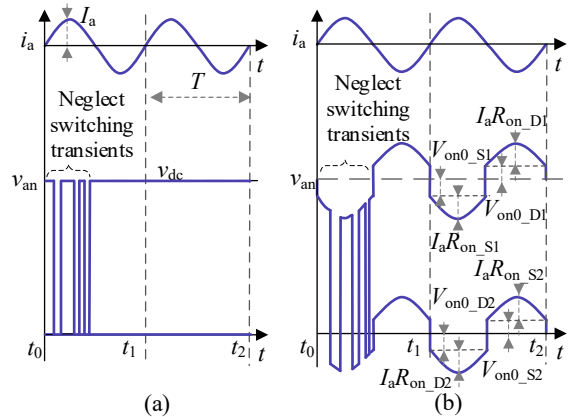

Fig. 4: On-state voltage of $S_1$, $S_2$, $D_1$, and $D_2$ adds errors to the inverter phase A output voltage $v_{an}$. (a) No on-state voltage in phase A. (b) On-state voltage in phase A are non-ideal.

The inverter output voltage waveform is not only affected by parasitic parameters of power semiconductor devices, but also depends on operating parameters, i.e., the sign of gate signals $v_{gate}$ and the sign of the phase current. For instance, in Fig. 4 (b), $S_1$ and $D_2$ conduct in a half fundamental cycle $T/2$ when the phase current $i_a$ is positive. The on-state voltage of $S_1$ reduces the upper envelope of $v_{an}$ from $v_{dc}$ to $v_{dc}$-$V_{on\_S1}$, while the on-state voltage of $D_2$ reduces the bottom envelope of $v_{an}$ from zero to 0-$V_{on\_D2}$.

### D. Assumptions in the Analysis

It is worth notifying that this paper analyzes the harmonics based on two assumptions: 1) Only resistive load is considered so the power factor is one which simplifies the analysis. 2) The voltage drop on the filter stage does not lead to a significant phase shift between the inverter output voltage ($v_{an}$, $v_{bn}$, $v_{cn}$) and the phase current $i_a$.

## II. Proposed Modelling Method of Harmonics in Control Variables $v_{dq}^*$

This section models the impact of the increased on-state voltage $V_{on}$ on the reference voltage harmonics both in the *abc* and the *dq* frame. More specifically, the resistive part $R_{on}$ is studied since an increased $R_{on}$ can potentially indicate bond wire fatigue.

The modelling scenario is the on-state resistance $\Delta R_{on}$ of one power semiconductor device $S_1$ reaches the end-of-life criteria, while on-state resistances of all the other power semiconductor devices have negligible variation. The proposed modelling method is also applicable for complicated degradation scenarios, e.g., multiple devices with different health statuses. A flowchart of the modelling method is shown in Fig. 5.

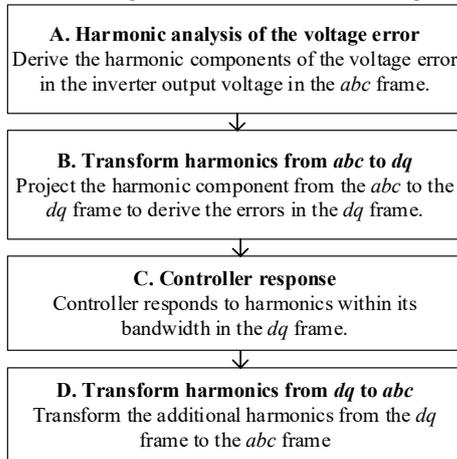

Fig. 5 Flow chart of harmonic calculation based on the increased on-state voltage error.

The method starts from analyzing the harmonic components in the voltage error in both *abc*- and *dq*-frame. The analyzed harmonic errors are used to study the controller response in both *dq*- and *abc*-frame. This section follows the flowchart to model harmonics in control variables when resistance $\Delta R_{on}$ of one power semiconductor device $S_1$ reaches the end-of-life criteria.

### A. Derive the Harmonics of Voltage Error in the *abc* Frame

The increased resistance $\Delta R_{on}$ adds errors $\Delta v_{an}$ to the phase A, as given in (3). The error $\Delta v_{an}$ also depends on the phase current magnitude $I_a$, the sign of the phase current $v_{ia+}$, and the sign of $S_1$ gate signal $v_{gate1+}$.

$$\Delta v_{an} = \Delta R_{on} I_a \sin(\omega t + \theta_g) v_{ia+} v_{gate1+},$$

$$v_{ia+} = \frac{1}{2} + \frac{2}{\pi} \sum_{k=1}^{\infty} \frac{\sin((2k-1)(\omega t + \theta_g))}{2k-1}, \quad (3)$$

$$v_{gate1+} = \frac{1}{2} + \frac{1}{2} m_d \sin(\theta_g)$$

where, $\omega$ is the grid voltage frequency, $t$ is time, $\theta_g$ is the phase angle of the grid voltage, $k$ is the harmonic order, and $m_d$ is the modulation index. The voltage error in phases B and C are zero because there are no additional resistances in these two phases.

$$\Delta v_{bn} = \Delta v_{cn} = 0 \quad (4)$$

### B. Analyze Errors in the *dq* Frame

The *dq* frame voltage error is defined as a column vector $\Delta v_{dq0}$=$[\Delta v_d\ \Delta v_q\ \Delta v_0]^T$ where $\Delta v_d$, $\Delta v_q$, $\Delta v_0$ are errors in the *dq0* axis, respectively. Equation (5) derives $\Delta v_{dq0}$ with the transformation matrix $M$ and the voltage errors in the *abc* frame $\Delta v_{abcn}$=$[\Delta v_{an}\ \Delta v_{bn}\ \Delta v_{cn}]^T$. The matrix $M$ can be found in [17].

$$\Delta v_{dq0} = M\Delta v_{abcn} = \frac{\Delta v_{an}}{3} \begin{bmatrix} 2\cos(\omega t) \\ -2\sin(\omega t) \\ 1 \end{bmatrix} \quad (5)$$

### C. Controller Response

The current controller responds to the voltage error in the *dq* frame, $\Delta v_{dq0}$, and adds harmonics to the reference voltage $\Delta v_{dq0}^*$=$[\Delta v_d^*\ \Delta v_q^*\ \Delta v_0^*]^T$ as shown in equation (6).

$$\Delta v_d^* \approx \Delta v_d,$$
$$\Delta v_q^* \approx \Delta v_q, \quad (6)$$
$$\Delta v_0^* = 0$$

where, $\Delta v_d^*$, $\Delta v_q^*$, and $\Delta v_0^*$ are additional harmonics in the reference voltages at the $dq0$ axis, respectively. The approximately equal sign means the controller can only respond to part of the errors within the control loop bandwidth. $\Delta v_0^*$ is zero because the zero-sequence error in equation (5) has no conduction path in the circuit.

### D. Controller Response in the *abc* Frame

The reference voltage in the *abc* frame $\Delta v_{abc}^* = [\Delta v_a^* \; \Delta v_b^* \; \Delta v_c^*]^T$ is derived in (7), where $M^{-1}$ is the inverse transformation matrix [17]. $\Delta v_a^*$, $\Delta v_b^*$, and $\Delta v_c^*$ are increased harmonics in the voltage references in the *abc* phase, respectively.

$$\begin{bmatrix} \Delta v_a^* \\ \Delta v_b^* \\ \Delta v_c^* \end{bmatrix} = M^{-1} \begin{bmatrix} \Delta v_d^* \\ \Delta v_q^* \\ 0 \end{bmatrix} \approx \frac{\Delta v_{an}}{3} \begin{bmatrix} 2 \\ -1 \\ -1 \end{bmatrix} \quad (7)$$

According to (7), the increased harmonics in reference voltage depend on the phase of the degraded component. Since degraded $S_1$ is in phase A, $\Delta v_a^*$ is in phase with $\Delta v_{an}$, while $\Delta v_b^*$ and $\Delta v_c^*$ are opposite to $\Delta v_{an}$.

## III. Analysis and Validation

### A. Simulation Model

To verify the equations listed in Section II, an inverter model is built in PLECS. Parameters of circuit components and control loops are listed in Table II. The output power is 5 kW. The initial phase angle $\theta_g$ of the phase voltage is $\pi/2$. The initial on-state voltage of IGBTs and diodes are set to $V_{on0}=0.75$ V and $R_{on0}=22.5$ m$\Omega$.

The increased resistance $\Delta R_{on}$ of $S_1$ is selected to be 5% of the initial resistance reaching the end-of-life criteria [5].

$$\Delta R_{on} = 5\% R_{on0} \approx 1 \; m\Omega \quad (15)$$

**Table II: Parameters of the inverter model in PLECS.**

| Parameter | Symbol | Value |
| --- | --- | --- |
| Output power | $P_o$ | 5 kW |
| DC link voltage | $V_{dc}$ | 800 V |
| Grid phase voltage amplitude | $V_g$ | 311 V |
| Bus capacitance | $C_{bus}$ | 600 μF |
| Bus capacitor ESR | $R_C$ | 1 mΩ |
| Filter inductance | $L_g$ | 6 mH |
| Filter inductor ESR | $R_L$ | 100 mΩ |
| Grid frequency | $f_g$ | 50 Hz |
| Switching frequency | $f_{sw}$ | 20 kHz |
| Sampling frequency | $f_{sa}$ | 20 kHz |
| Sampling period | $T_{sa}$ | 50 μs |
| Voltage loop PI parameters | $K_{pv}/K_{iv}$ | 0.6/13 |
| Voltage loop bandwidth | $f_{Bv}$ | 92 Hz |
| Current loop PI parameters | $K_{pc}/K_{ic}$ | 40/500 |
| Current loop bandwidth | $f_{BI}$ | 1.2 kHz |
| Deadtime | $t_{deadtime}$ | 1 μs |
| The initial on-state voltage part of all IGBTs and diodes | $V_{on0}$ | 0.75 V |
| The initial resistive part of all IGBTs and diodes | $R_{on0}$ | 22.5 mΩ |
| Increased on-state resistance of $S_1$ | $\Delta R_{on}$ | 1 mΩ |

### B. Result

Comparisons between equations and simulations are shown in Fig. 6. Basically, the harmonic distribution follows equations (3)-(7). The amplitudes of the harmonics reduce as the harmonic order increases.

The *d*-axis harmonics $\Delta V_d^*$ are slightly higher than the *q*-axis harmonics $\Delta V_q^*$. Since these two harmonics are in the range of several mV, using the higher harmonics $\Delta V_d^*$ can have lower errors. The increased harmonics in phase A voltage reference $\Delta V_a^*$ are two times of the harmonics in phases B and C voltage references $\Delta V_b^*$ and $\Delta V_c^*$, because the degraded component $S_1$ is in phase A. According to equation (7), the phase angles of $\Delta V_b^*$ and $\Delta V_c^*$ are opposite to the $\Delta V_a^*$. The harmonics vectors are shown in Fig. 7. Fig. 7 also shows that the total harmonics of each phase are the vector sum of the original harmonics ($V_a^*$, $V_b^*$, and $V_c^*$) and the increased harmonics ($\Delta V_a^*$, $\Delta V_b^*$, and $\Delta V_c^*$). Therefore, a lower level of original harmonics helps to identify the increased harmonics due to increased on-state resistances of power semiconductor devices.

To further verify the model, more scenarios with different resistances are simulated: $\Delta R_{on}$ varies between 0 and 1 mΩ with a 0.1 mΩ step size. The result of $|\Delta V_d^*|$ and $|\Delta V_q^*|$ are shown in Fig. 8. The simulation results (circles) are falling on the equation-based results (solid lines) with minor errors. The average relative errors of each harmonic order are listed in Table III. The relative error between equations and simulations increases as the frequency of the harmonics increases. Therefore, low-order harmonics are more suitable for the on-state voltage estimation. For instance, the relative error is around 0.44% using the DC-component at the *d* axis to estimate the $\Delta R_{on}$, which is lower than the higher-order $d$-axis harmonics-based estimation errors.

One of the reasons for the increasing error at the high-frequency region is that the loop gain reduces at higher frequencies. Fig. 9 shows the effect of the control loop bandwidth. It compares the transfer function from the voltage error $V_{on}$ to the reference voltage $v_{dq}^*$ and a frequency sweep results from simulation. The loop gains, phase shifts, and relative errors are listed in Table IV. The loop gain reduces from one to 0.77 and the phase shift reduces from zero to around -53° as the frequency increases from zero to the 24th order of the fundamental. The changing phase shift also limits the harmonic suppression capability because voltage errors are not fully cancelled. Therefore, the error increases from 0% to around 83% as the frequency increases from DC to the 24th-order harmonics.

The sensitivity is also a main challenge in condition monitoring techniques since the health indicator only shift in a small range. Thus, the sensitivity of control variable harmonics to the increased resistance $\Delta R_{on}$ is listed in Table V. The phase current amplitude $I_a$ is normalized to 1 A and the modulation index $m_d$=0.775. The sensitivity reduces gradually as the frequency increases which also leads to the same conclusion that low-order harmonics should be used to monitor the health status. Table V also shows that the voltage harmonics in the $dq$ frame control variables will increase when the phase current increases.

### C. Limitation

There are two main limitations of the modelling method.
- The increased resistance is only one of the harmonic sources. The shift in on-state resistance is in the level of several mΩ over the whole lifespan resulting in harmonic increments at the level lower than mV. The harmonics can easily be covered by the other harmonic sources, e.g., the grid voltage harmonics and the inverter current harmonics. Further study is needed to quantify the effect of other harmonics.
- The inverter output current is sampled, filtered, and analog-digital converted. These processes should have a high resolution to convey the voltage errors into the current control loop.

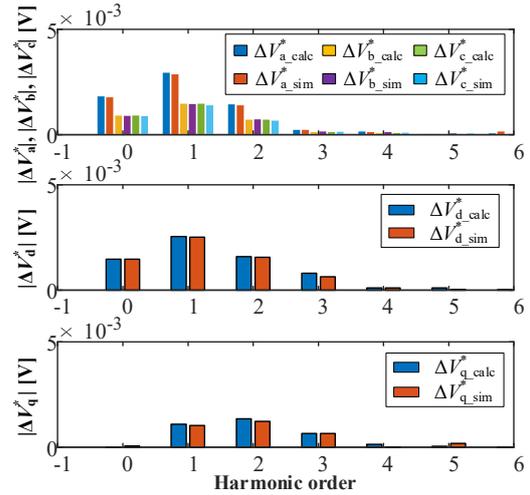

Fig. 6: Comparison of harmonics magnitude between equations (3)-(7) and simulation. The increased on-state resistance of $S_1$ is $\Delta R_{on}$=1 mΩ.

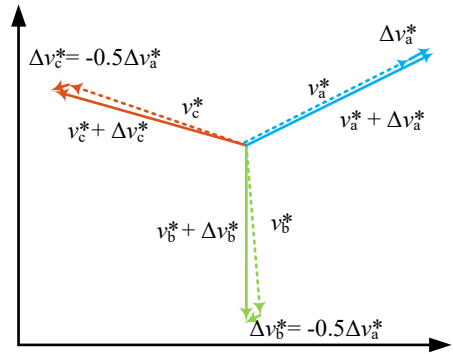

Fig. 7: Vectors of the increased harmonics in the $abc$ frame. Increased harmonics in the phases B and C ($\Delta V_b^*$ and $\Delta V_c^*$) are opposite to the increased harmonic in phase A ($\Delta V_a^*$).

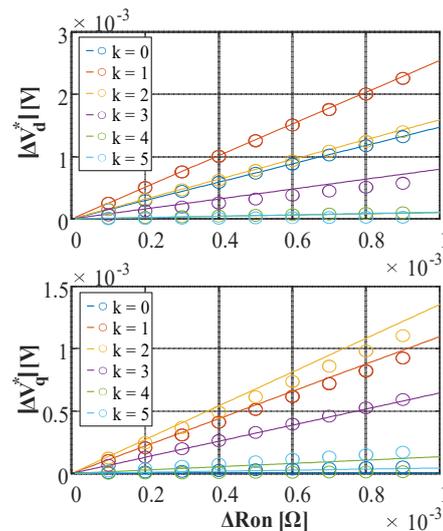

Fig. 8: Comparison of harmonics between equations (3)-(7) (solid lines) and simulations (circles). $\Delta R_{on}$ of $S_1$ varies from 0 to 1 mΩ, and the harmonic order $k$ varies from 0 to 5.

**Table III: Average relative error between equation (3)-(7) and simulation of $V^*_{dq}$.**

| Order | 0 | 1 | 2 | 3 | 4 | 5 |
|---|---|---|---|---|---|---|
| $\Delta v^*_{derr}$ (%) | 0.44 | 1.27 | 2.29 | 20.24 | 9.00 | 71.49 |
| $\Delta v^*_{qerr}$ (%) | ~ | 6.32 | 9.30 | 1.95 | 88.51 | 336.93 |

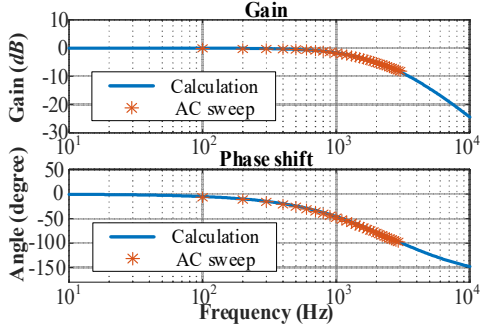

Fig. 9: Bode plot of transfer function from the voltage error $V_{on}$ to the voltage reference $v^*_{dq}$.

**Table IV: Closed-loop gain of the on-state voltage $V_{on}$ to voltage reference $v^*_{dq}$.**

| Order | 0 | 6 | 12 | 18 | 24 |
|---|---|---|---|---|---|
| $|G_{Von}|$ | 1 | 0.98 | 0.93 | 0.85 | 0.77 |
| Phase (°) | 0 | -15 | -29 | -42 | -53 |
| \|Error\| (%) | 0 | 25.48 | 48.64 | 67.92 | 82.80 |

**Table V: Sensitivity of control variable harmonics on the $\Delta R_{on}$.**

| Order | Sensitivity of the $d$-axis reference (V/Ω) | Sensitivity of the $q$-axis reference (V/Ω) |
|---|---|---|
| 0 | 0.1382 | 0.0000 |
| 1 | 0.2383 | 0.1030 |
| 2 | 0.1491 | 0.1272 |
| 3 | 0.0747 | 0.0606 |
| 4 | 0.0094 | 0.0125 |
| 5 | 0.0101 | 0.0040 |
| 6 | 0.0010 | 0.0021 |

## Conclusion

This paper analyzes the mechanism of how the on-state voltage $V_{on}$ induces harmonics in the current control loop of a voltage source inverter. Since the harmonics can be observed from different variables in the control loop, these variables can be used to estimate $V_{on}$ and the equivalent resistance $R_{on}$. Using the current controller output voltage $v^*_d$ and $v^*_q$ has an advantage of monitoring three-phase components with only two variables.

The on-state voltage increment is divided into two parts ($\Delta V_{on}=\Delta V_{on0}+\Delta R_{on}I$). The resistive part $\Delta R_{on}$ is studied since a shifting $\Delta R_{on}$ can indicate failure mechanisms, e.g., bondwire fatigue. The harmonics depend on the modulation index $m_d$, the phase current amplitude, sign of the gate signal, and sign of the phase current. These dependent variables are already used for control purposes so there is no need for additional sensing devices.

Model performances are analyzed and compared with PLECS simulation. The model can predict the harmonic values with some limitations. When the frequency increases, the sensitivity reduces and the estimation error increases. Thus, low-frequency harmonics are feasible to estimate $\Delta V_{on}$. $\Delta R_{on}$ estimations based on the $d$-axis DC component have errors below 0.5%. One of the limitations is that the on-state voltage $V_{on}$ is not the only source of harmonics. Future study should focus on the effects of the other harmonics sources as well. Another limitation is the current sensor should have a high resolution to observe the voltage errors.

## References

[1] A. Golnas, "PV System Reliability: An Operator's Perspective", *IEEE J. Photovol.*, vol. 3, no. 1, pp. 416-421, Jan. 2013.

[2] M. Wilkinson, K. Harman, F. Spinato, B. Hendriks and T. Van Delft, "Measuring wind turbine reliability - results of the reliawind project", *European Wind Energy Conference*, Mar. 14–17, 2011.

[3] J. M. Freeman, G. T. Klise, A. Walker, and O. Lavrova, "Evaluating energy impacts and costs from PV component failures," in *Proc. IEEE 7th World Conf. Photovolt. Energy Convers.*, Jun. 2018, pp. 1761–1765.

[4] M. S. Haque, S. Choi and J. Baek, "Auxiliary Particle Filtering-Based Estimation of Remaining Useful Life of IGBT," *IEEE Trans. Ind. Electron.*, vol. 65, no. 3, pp. 2693-2703, March 2018.

[5] U. -M. Choi, F. Blaabjerg, S. Jørgensen, S. Munk-Nielsen and B. Rannestad, "Reliability Improvement of Power Converters by Means of Condition Monitoring of IGBT Modules," *IEEE Trans. Power Electron.*, vol. 32, no. 10, pp. 7990-7997, Oct. 2017.

[6] N. Valentine, "Failure modes and mechanisms analysis of silicon power devices," MSc. Thesis, Dept. of Mech. Eng., Univ. of Maryland, College Park, 2017.

[7] H. Oh, B. Han, P. McCluskey, C. Han and B. D. Youn, "Physics-of-Failure, Condition Monitoring, and Prognostics of Insulated Gate Bipolar Transistor Modules: A Review," *IEEE Trans. Power Electron.*, vol. 30, no. 5, pp. 2413-2426, May 2015.

[8] Y. Peng, Y. Shen and H. Wang, "A Converter-Level on-State Voltage Measurement Method for Power Semiconductor Devices," *IEEE Trans. Power Electron.*, vol. 36, no. 2, pp. 1220-1224, Feb. 2021.

[9] V. Smet, F. Forest, J. -J. Huselstein, A. Rashed and F. Richardeau, "Evaluation of Vce Monitoring as a Real-Time Method to Estimate Aging of Bond Wire-IGBT Modules Stressed by Power Cycling," *IEEE Trans. Ind. Electron.*, vol. 60, no. 7, pp. 2760-2770, Jul. 2013.

[10] C. Roy, N. Kim, D. Evans, A. P. Sirat, J. Gafford and B. Parkhideh, "A Half-Bridge On-State Voltage Sensor for In-Situ Measurements," *Proc. IEEE Energy Convers. Congr. Expo.*, 2022, pp. 1-7.

[11] S. H. Ali, X. Li, A. S. Kamath and B. Akin, "A Simple Plug-In Circuit for IGBT Gate Drivers to Monitor Device Aging: Toward Smart Gate Drivers," *IEEE Power Electron. Mag.*, vol. 5, no. 3, pp. 45-55, Sep. 2018.

[12] D. Xiang, L. Ran, P. Tavner, S. Yang, A. Bryant and P. Mawby, "Condition Monitoring Power Module Solder Fatigue Using Inverter Harmonic Identification," *IEEE Trans. Power Electron.*, vol. 27, no. 1, pp. 235-247, Jan. 2012.

[13] F. Yüce and M. Hiller, "Investigation of Bond Wire Lift-Off by Analyzing the Controller Output Voltage Harmonics for the Purpose of Condition Monitoring," in *Proc. 22nd Eur. Conf. Power Electron. Appl. (EPE ECCE Europe)*, pp. P.1-P.10, Sep. 2020.

[14] F. Yüce and M. Hiller, "Condition Monitoring of Power Electronic Systems Through Data Analysis of Measurement Signals and Control Output Variables," *IEEE Trans. Emerg. Sel. Topics Power Electron.*, vol. 10, no. 5, pp. 5118-5131, Oct. 2022.

[15] A. Sangwongwanich, A. Abdelhakim, Y. Yang, and K. Zhou, "Control of single-phase and three-phase DC/AC converters," in *Control of Power Electronic Converters and Systems*. F. Blaabjerg, Ed. Cambridge: Academic Press, 2018, pp. 153–173.

[16] S. Moon, "Hybrid PWM update method for time delay compensation in current control loop," Ph.D. diss., Dept. Elect. Eng., Virginia Polytechnic Inst. State Univ., Blacksburg, VA, USA, 2017.

[17] Q.-C. Zhong and T. Hornik, *Control of Power Inverters in Renewable Energy and Smart Grid Integration*, vol. 97. Hoboken, NJ, USA: Wiley, 2012.